\documentclass[11pt]{article}

\usepackage[T1]{fontenc} 
\usepackage[utf8]{inputenc}
\usepackage{amsfonts} 
\usepackage{amssymb} 
\usepackage{amsmath} 
\usepackage{graphicx} 
\usepackage{subfigure} 
\usepackage{array} 
\usepackage{dcolumn} 
\usepackage{bm} 
\usepackage{latexsym} 
\usepackage{longtable} 
\usepackage{bbold}
\usepackage[dvipsnames]{xcolor}
\usepackage{multirow}
\usepackage{rotating}
\usepackage{comment}
\usepackage{bm}
\usepackage{placeins}
\usepackage{lipsum}
\usepackage[symbol]{footmisc}
\usepackage{hyperref}
\usepackage{fullpage}
\usepackage{authblk}

\graphicspath{{./figs/}}
\allowdisplaybreaks
\nopagebreak

\include{macro}

\newcommand{\citet}{\cite}

\begin{document}

\renewcommand{\thefootnote}{\fnsymbol{footnote}}


\title{AI and Theoretical Particle Physics}

\author[1\footnote{rajan@lanl.gov}]{Rajan Gupta}
\author[1\footnote{tanmoy@lanl.gov}]{Tanmoy Bhattacharya}
\author[2\footnote{boram@lanl.gov}]{Boram Yoon}
\affil[1]{T-2, Theoretical Division, Los Alamos National Laboratory, Los Alamos, NM 87545, USA}
\affil[2]{CCS-7, Computer, Computational, and Statistical Sciences division, Los Alamos National Laboratory, Los Alamos, NM 87545, USA}
\date{}
\maketitle


%
%
%
%
%
\begin{abstract}
Theoretical particle physicists continue to push the envelope in both high performance computing and in managing and analyzing large data sets. For example, the goals of sub-percent accuracy in predictions of quantum chromodynamics (QCD) using large scale simulations of lattice QCD and in finding signals of rare events and new physics in exabytes of data produced by  experiments at the high luminosity large hadron collider (LHC) require new tools beyond just developments in hardware. Machine learning and artificial intelligence offer the promise of dramatically reducing the computational cost and time. This chapter reviews selected areas where AI/ML tools could have a major impact, provides an overview of the challenges, and discusses how new ideas such as normalizing flows can speed up the generation of gauge configurations needed in lattice QCD calculations; the growth of ML in  surrogate models and pattern matching to reduce the cost of event generators and in the analysis of experimental data; and in the search for viable vacua in the landscape of string theories. While such approaches transform aspects of particle theory into computational problems, and thus black boxes, we argue that physics-aware development of these tools combined with algorithms that ensure that the results are bias free will continue to require a deep understanding of the physics. We see this broader transformation as akin to formulating and extracting observables from simulations of lattice QCD, a numerical integration of the path integral formulation of QCD that nevertheless requires a deep understanding of the underlying quantum field theory, the standard model of particle physics and effective field theory methods. \looseness-1

\end{abstract}
%

\newpage 
\tableofcontents
\newpage 
\section{Introduction}
\label{sec:intro}

Modern high-energy physics analyses make extensive use of large-scale computing, and many advances in computing were driven by the needs of this community. The needed computational resources are, however, fast outpacing the growth in hardware capability, therefore new techniques and algorithms are  needed to reach the design goals of precision experiments, and to make predictions of the Standard Model (SM) and theories beyond in order to  compare them to experiments. Machine learning (ML) provides a very important tool in this respect: it provides a general approach for devising algorithms to approximately solve a multitude of problems at far less cost---both for code development and for execution---than traditional methods.

One can view machine learning as a data-driven development of pattern matching, interpolation schemes or surrogate models.  Most of the powerful machine-learning systems available today are black boxes, whose detailed analytical structure is difficult, if not impossible, to elucidate. This, however, is nothing new to computational high-energy physicists: numerical methods like lattice quantum chromodynamics (QCD)~\cite{Creutz:1984mg,Gattringer:2010zz} have long supplanted analytical calculations even though the simulations are inscrutable (see Sec.~\ref{sec:bb}). The only true requirement, as always, is not an analytic understanding, but correctness guarantees assured by a principled estimation of bias and uncertainty in the predictions.  In this chapter, we discuss the application of machine learning in speeding up predictions of the SM  and beyond the standard model (BSM) theories, speeding up simulations of lattice QCD, in fitting data including global fits to obtain parton distribution functions (PDFs), and constraining the space of possible BSM and string theories. The scope  of ongoing research is large and the pace frantic. We aim to communicate some of the excitement in this review.   

 
\section{Surrogate models: looking for a needle in a haystack}
\label{sec:surrogate}

Three very high impact and related uses of ML in high energy physics (HEP) are to (i) connect individual detector outputs from billions of events (petabytes of experimental  data) to an [un]characterized  
signal, (ii) to provide 
signals that distinguish different BSM theories as a 
function of the free parameters 
that characterize these theories, and (iii) survey large regions of possible [string] theories that are not ruled out by known low energy physics.  Machine learning is very well suited for  such pattern matching and searches to find a signal (well-characterized or hidden) from a
large body of data. It is typically a data-driven process 
without an underlying theory, but symmetries and constraints of the theory can/should be built in to make it more physics-aware. 
Below we give examples of how ML systems have transformed such analyses. 

\subsection{Classifying an event}
\label{sec:event}
A typical event in collider experiments consists of hits (energy or charge deposition) in thousands of detector units. These hits (called particle-level events) are caused by standard model particles that survive the evolution between the initial collision vertex and the final detection, or are created in between as illustrated in Fig.~\ref{fig:ttbarEvent}. What we are interested in are signatures of new physics. This requires reconstructing the  vertex since novel particles are expected to survive for a very short time, i.e., in a millimeter or smaller sized region about the vertex. This involves deciphering from a large volume of high dimensional detected data  (labeled particle-level) the [virtual, exotic] particles produced right after the initial collision (labeled parton level), and their decay channels. The straightforward first use of ML to reduce dimensionality is to show it millions of such events and train it to classify patterns. While this is doable, it is inefficient. SM theory tells us that charged energetic SM particles produce jets (collimated beams of particles typically coming from a single progenitor) and building in this knowledge is essential for efficiency. For example, for strong interactions described by QCD, one could classify jets by whether they arose from a quark or a gluon progenitor. Thus, one could first train a ML system to recognize quark and gluon jets and how to distinguish between them and then classify each particle-level event (hits in thousands of detectors) in terms of jets and their progenitors. This reduction in dimension (data compression) from detector signals to jets to a collision (called particle-level to parton-level to event-level) is the sought for translation between signals in detectors and the language of physics~\cite{Kasieczka:2019dbj,Faucett:2020vbu,Bradshaw:2022qev}. 

ML systems are ideally suited for this task and are now ubiquitous and essential for such analyses (see David Rousseau, {\it ibid}). Examples of recent developments are~\citet{Komiske:2022enw} for extracting information on jet substructure, and  SymmetryGAN~\citet{Desai:2021wbb} for inferring symmetries.

\subsection{Scattering amplitudes and cross-sections}
\label{sec:amp}
\begin{figure}[ht] 
\subfigure
{
    \includegraphics[width=0.47\linewidth]{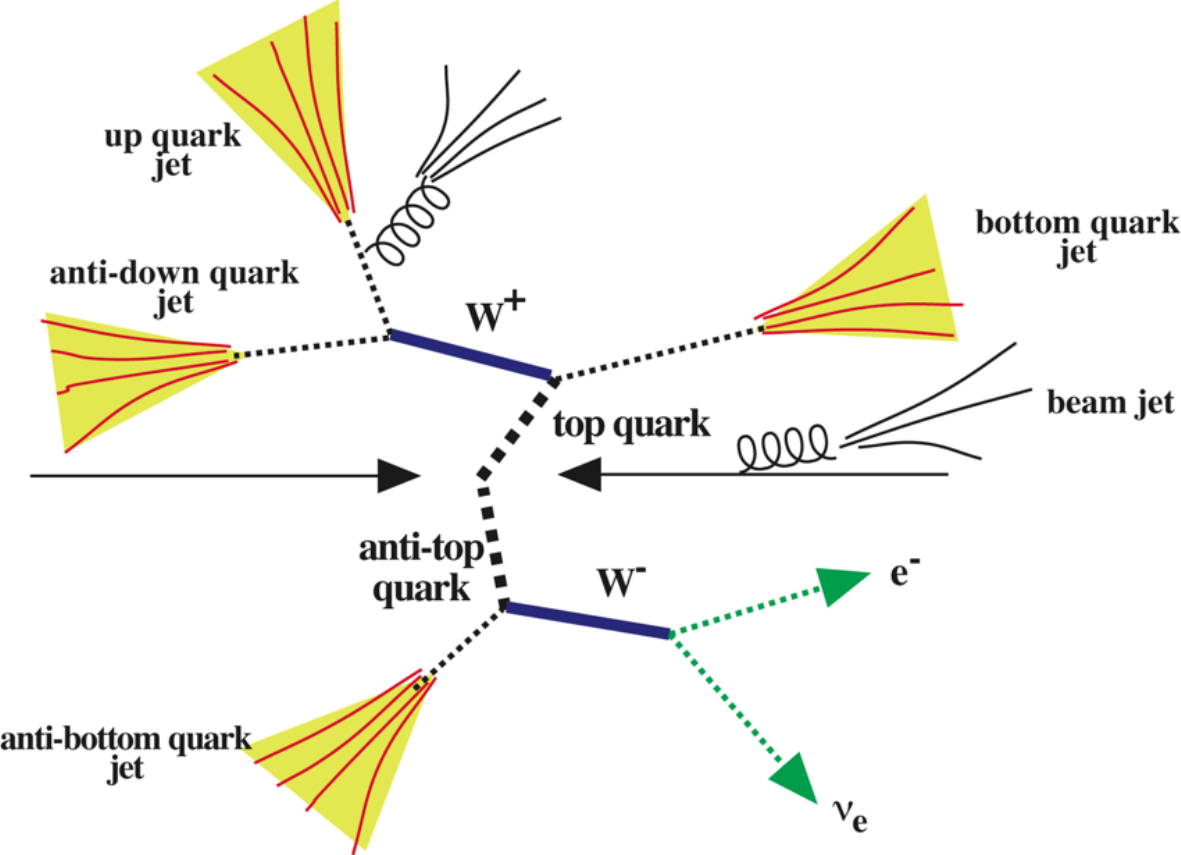}  
    \includegraphics[width=0.47\linewidth]{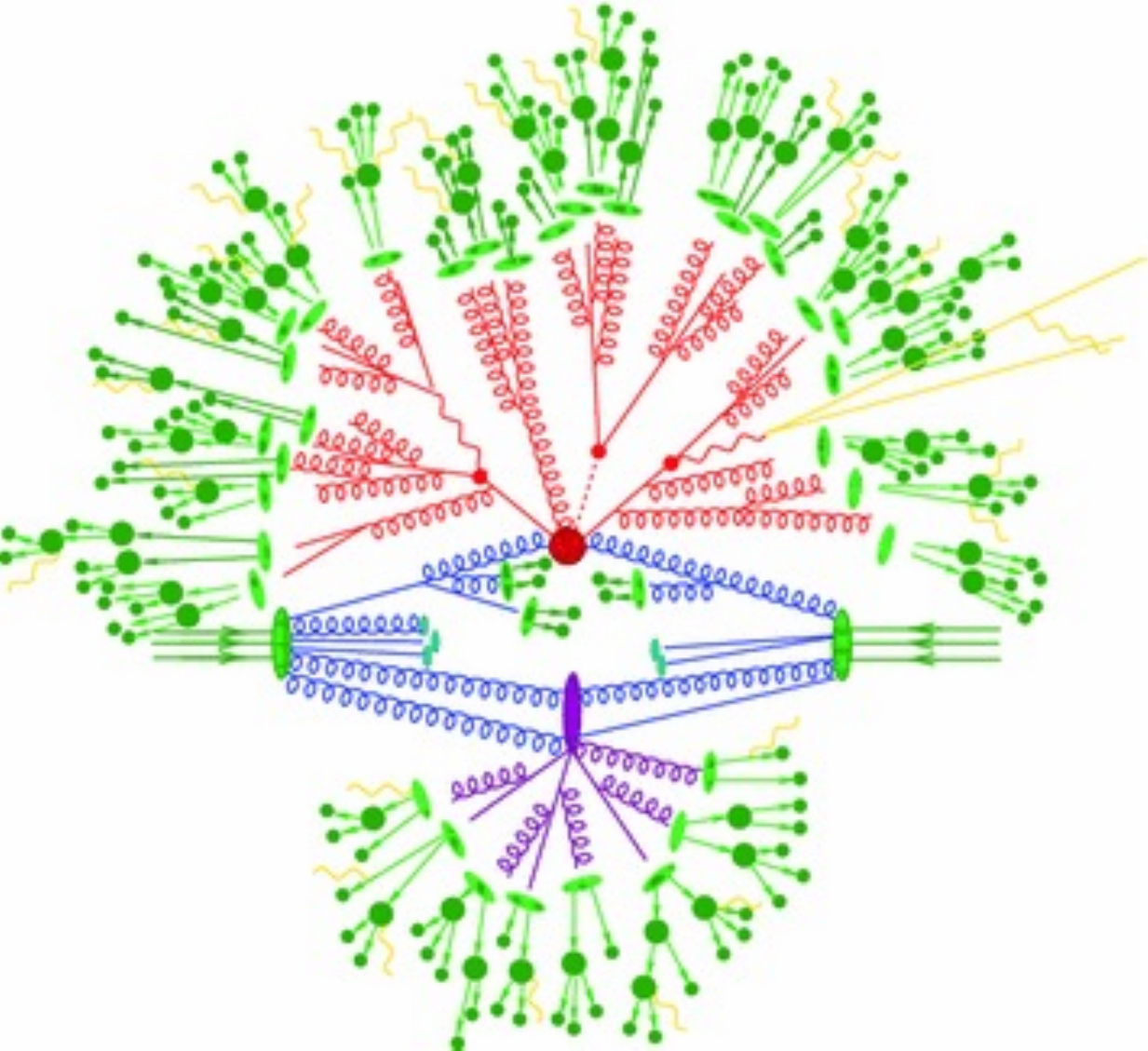}
    }
\caption{(Left) An event initiated by electron-antielectron (or quark-antiquark) annihilation  into a $t,\bar t$ quark pair production. The $t$ and $\bar t$ are very short lived ($5\times 10^{-25}$ sec) and decay into $W^- b\  W^+ \bar b $. The $W$ boson and the $b$ quarks also decay rapidly ($3\times 10^{-25}$ and $\sim 10^{-12}$ secs, respectively) producing bottom, up and antidown quark ($b, \bar b, u, \bar d$) jets.  Also shown are two other jets: one from the radiation of an energetic photon from the initial electron and one from gluon emission from the up quark.  In this event, the final state has six partons ($b, \bar b, u, \bar d, g$ and $\gamma$) that produce jets, and the neutrino ($\nu_e$) and $e^-$ from $W^-$ decay that fly away. (Figure produced by Pekka Sinervo). (Right) A $p p$ event with hard collision between two constituent partons (red circle) that creates parton jets (red). Collisions between remnants of $p p$ (purple) similarly produce away side partons. These partons transform into jets of hadrons (green) that trigger the detectors. Soft gluons that neutralize color are not shown in either figure.
\label{fig:ttbarEvent}
}
\end{figure}

Theoretical predictions of expected outcomes in high-energy collider experiments are differential cross-sections. These require the calculation of quantum mechanical amplitudes for a certain well-specified initial state going into a given final state of partons (amplitude for parton-level description) as illustrated in Fig.~\ref{fig:ttbarEvent}. These partons are then transformed into particles that reach the detectors using jet generators. 

For higher precision and greater detail, theorists calculate differential cross-sections with increasingly more partons in the final state and with fewer phase space variables (4-momenta of various particles) integrated over. These corrections to the basic interaction are ordered in terms of details: leading order (LO), next-to-leading order (NLO) and so forth. Attaining each order of these theoretical calculations has required years of effort by the community. These amplitudes, say at next-to-next-to-leading order (NNLO), involve multidimensional integrals over the allowed phase space (momentum) of each virtual particle. The integrands are convolutions and complex; they can have large logarithms, singular regions, large cancellations and highly peaked structure that require care during numerical integration. Physicists know how to deal with these and get finite physical answers, however, the evaluation is very computationally expensive (up to millions of core hours per event).

The output, on the other hand, is a well-behaved number (distribution). Furthermore, these results are governed by symmetries such as invariance under rotation about the initial collision axis, that have to be implemented in the analysis. The goal is to replace these expensive calculations with ML regression models. A typical procedure for generating parton-level amplitudes is to (1) choose random samples of phase space points for a given initial state in a known theory (say the standard model) and evaluate the amplitudes or cross-sections using  traditional approaches, (2) train an ML regression model to predict the same quantities for each phase space point, and (3) use the trained ML to predict amplitudes or cross-sections for new desired points, instead of the computationally expensive traditional calculations. In this procedure, the ML regression model learns about the correlation between the input and output variables and efficiently interpolates and extrapolates from the input parameters (or output)  used for the training data for new evaluation points. 

The reduction in the computation cost due to not having to do the complex integrals can be factors of $10^3$--$10^6$ or even more. These data can then be compared to experimental data (similarly compressed) to look for signals (agreements, disagreements, or  anomalies) via pattern matching. 

In \citet{Bishara:2019iwh}, the authors trained a Gradient boosting decision tree (GBDT) regression algorithm to make predictions of the amplitudes of the $gg\rightarrow ZZ$ process for a given phase space point. A similar approach approximating the multi-variable scattering amplitudes of $e^+ e^- \rightarrow \hbox{5-jets}$ using a fully-connected neural network has been explored in \citet{Badger:2020uow}.  

To constrain the parameters of a BSM theory, this process can be carried out for various values of couplings and masses and a further ML is used to interpolate over possible values. 
In \citet{Buckley:2020bxg}, the authors trained a distributed Gaussian processes (DGP) regression algorithm to predict the Minimal Supersymmetric SM (MSSM) cross-sections at NLO.

A key issue in using such surrogate models, or any numerical calculation in general, is uncertainty estimation. This is important in many fields where the results of ML form the basis of action with serious consequences for incorrect decisions. But, in scientific disciplines as well, the same control over uncertainties is needed for the field to make progress.  The purpose of theoretical calculations such as these is often to provide, as a gold standard, predictions of the theory against which other approximate techniques or experimental evidence can be evaluated. Not having a robust estimate of the uncertainty in the gold standard dilutes its use, except when they are known to be negligible compared to other uncertainties in the calculation.

The problem with black box ML methods that we focus on is that they are often not amenable to an analysis of systematic errors. To counter this, they are often used in situations where they form an inner component of a self-correcting calculation: conceptually, they generate hypotheses whose correctness affects the computational load but not the accuracy of the answers. An example of such use is in  constructing proposals for an update of a gauge configuration in a Metropolis accept/reject step as discussed in Sec.~\ref{sec:lqcd}. For a general method for bias correction for averages of a surrogate observable, see Sec.~\ref{sec:bias}. 

\section{ML models as a fitting function}
In some analyses of physics data, the exact fitting functional form describing the underlying physics theory is not known. For such cases, ML regression models can be used to parameterize the physics data. The NNPDF collaboration~(http://nnpdf.mi.infn.it/)  proposed an approach for the global analyses of parton distribution functions (PDFs) using machine learning, called the NNPDF framework~\cite{Carrazza:2019mzf,Ball:2021leu,NNPDF:2021uiq}. PDFs describe the probability density of quarks and gluons in a hadron as a function of the momentum fraction $x$ carried by the parton. They cannot be computed directly, but are extracted from data using a very systematic combined analysis of theoretical predictions and experimental results, initially from HERA (Hamburg, Germany), and now from the large hadron collider (LHC) at Geneva, Switzerland, etc. The output is a Monte Carlo representation of PDFs and their uncertainties: a probability distribution in a space of functions. These PDFs are essential inputs in the search for new physics in such experiments.

As an example of the analysis, a neural network is used to parameterize the PDFs $f_k(x, Q_0)$ as
\begin{align}
    x f_k(x, Q_0) = A_k x^{1-\alpha_k} (1-x)^{\beta_k} \textrm{NN}_k(x),
    \label{eq:pdf}
\end{align}
where $k$ is the index of the eight different PDF bases, which are linear combinations of the quarks, antiquarks, and gluons, $Q_0$ is the parametrization scale, and $\textrm{NN}_k(x)$ is the neural network regression model. The NNPDF framework generates artificial Monte Carlo data replicas based on the experimental covariance matrix of each experiment, and determines the free parameters  $A_k$, $\alpha_k$, $\beta_k$, and the neural network parameters. To avoid the  issue of the scale of the momentum fraction $x$, the regression model takes two inputs variables of $(x, \ln x)$, and to minimize the over-fitting risk, they use a relatively small fully-connected neural network of 2-25-20-8 neurons, where the first and last numbers indicate the number of input and output nodes, and the two middle numbers indicate the size of the hidden layers. In the latest framework, NNPDF4.0, a single neural network with 8 output nodes is successfully trained for all 8 different PDF bases indexed by $k$.

In another area, to describe the  potential between two heavy quarks,~\citet{Shi:2021qri} used an artificial neural network as a general fitting functional.  They trained two artificial neural networks to represent the real and imaginary parts of the heavy quark potential using the lattice QCD data, and obtained a good model-independent empirical description of the heavy quark potential.

 \section{Multidimensional integration using machine learning}
\label{sec:integ}
Many physics calculations involve multidimensional integration whose analytic solution is not known. Numerical integration methods are used to evaluate the integrals, and are often  computationally expensive and bottlenecks in the overall analysis. 
When the dimension of the integration is large, as in lattice QCD (LQCD), Monte Carlo methods are the most efficient. Monte Carlo integration has two steps---efficient but random sampling of the space of independent variables and the  evaluation of the integrand for each random sample; ML can speed up both steps.

In most Monte Carlo integration algorithms implementing importance sampling, the goal of the random sampling is to draw samples from a specific distribution, such as the absolute value of the integrand whose value is accessible only through numerical evaluation for a given sample point. Various approaches, such as the VEGAS~\cite{PETERLEPAGE1978192, Lepage:123074, Lepage:2020tgj} and hybrid Monte Carlo (HMC)~\cite{Duane:1987de} algorithms, have been developed, and recently, new algorithms based on generative ML models have been proposed~\cite{Gao:2020vdv,Pawlowski:2018qxs,Albergo:2019eim,Kanwar:2020xzo,Hackett:2021idh}.

Reference \citet{Gao:2020vdv} provided a general numerical integration algorithm using ML and published an integration library: \textit{i-flow}. They implemented the importance sampling using normalizing flows~\cite{rezende2016variational,10.2307/2958830}, which is a family of generative models for efficient sampling and density evaluation of a specified probability distribution, and demonstrated its efficiency on various problems. 
For each sample, the integrand is calculated and the intergral  is estimated as its expectation value. 

For multi-dimensional integration, recognizing the efficient interpolation ability of ML models, \citet{Yoon:2020zmb} proposed training a regression model to emulate the integrand  using the collected samples and obtain a better estimate of the integral. In this approach, the effect of the difference between the true integrand and the ML model is quantified using the bias correction method described in Section~\ref{sec:bias}.

Lattice QCD calculations typically require integration over the gauge fields and the integration dimension in state-of-the-art calculations is $\mathcal{O}(10^{10})$. Markov Chain Monte Carlo (MCMC) methods are used to draw the samples, but are computationally expensive and produce samples with large autocorrelations. New approaches using generative ML models are under active development~\cite{Albergo:2022qfi,Pawlowski:2018qxs,Albergo:2019eim,Kanwar:2020xzo,Hackett:2021idh}. A more detailed discussion of lattice QCD is given in Section~\ref{sec:GG}.

\section{Lattice QCD}
\label{sec:lqcd}

Quantum chromodynamics (QCD)~\cite{Campbell:2017hsr,Ioffe:2010zz} is the fundamental theory of nature 
describing the strong interactions between quarks and gluons. Together 
with weak and electromagnetic interactions, it constitutes the standard model 
of elementary particles and their interactions~\cite{Langacker:2017uah}. The coupling constant of QCD, $\alpha_s(q^2)$, 
depends on the energy scale, $q^2$, of the interactions. For all natural phenomena 
characterized by $q^2 < 1$~GeV${}^2$, it is large, $O(1)$, consequently, standard methods based 
on perturbation theory (expansion in a small parameter) that are highly successful for 
weak and electromagnetic interactions are unreliable. In 1974, Wilson 
formulated QCD on 
a 4-D Euclidean lattice that provided a first principle method for solving the 
theory using numerical evaluations of its path-integral formulation~\cite{Wilson:1974sk,Wilson:1975id,Wilson:1979wp}. Physical results are obtained by taking the infinite volume and continuum limits of the data. Today, large-scale simulations of 
lattice QCD (LQCD) provide many high precision results that test QCD, provide input 
needed in the analysis of experiments and give  a detailed description of hadrons, the bound states of quarks and gluons~\cite{Aoki:2021kgd}.

\begin{figure}[tb]
    \centering
    \includegraphics[width=0.4\textwidth]{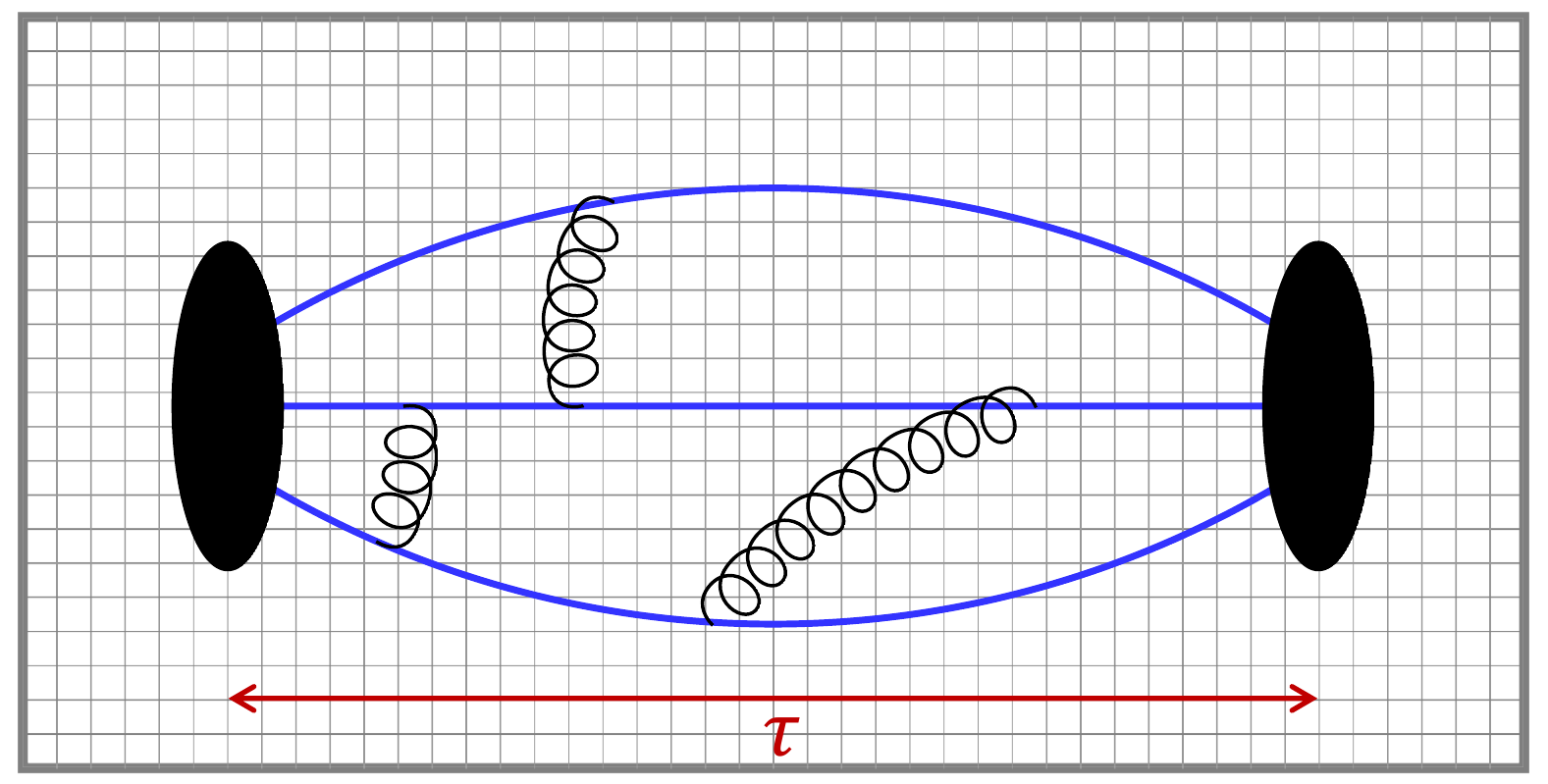}\qquad
    \includegraphics[width=0.4\textwidth]{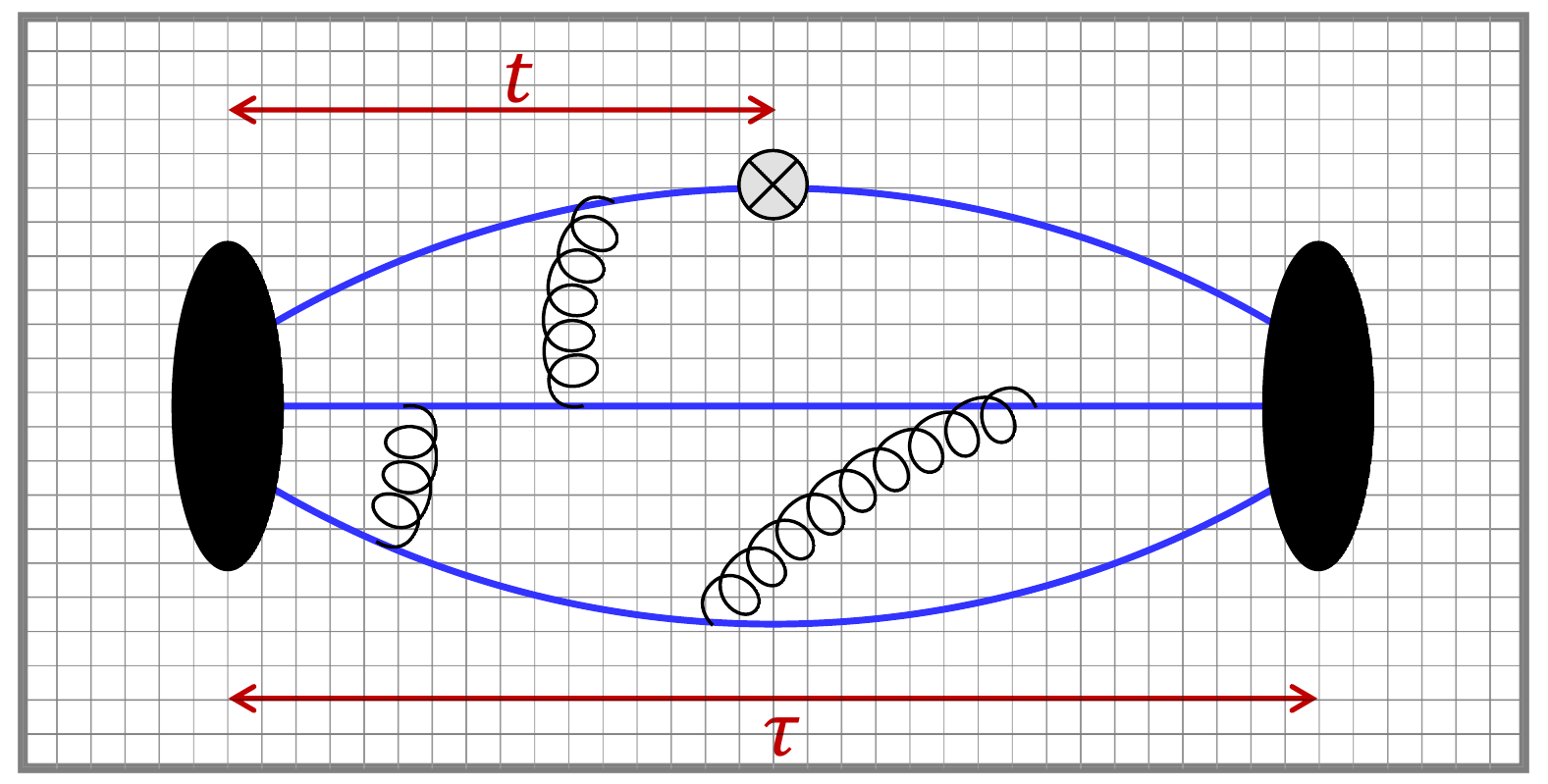}
    \caption{Quark-line diagrams of the two (left) and three-point (right) functions of the proton.  The black blobs denote nucleon source and sink,   separated by Euclidean time $\tau$.  The insertion of the operator in the 3-point function (labeled by $\otimes$) is done at time slice $t$ to calculate its matrix element within   the nucleon state.}
    \label{fig:lat_diags}
\end{figure}

Classical numerical simulations of LQCD generate gauge configurations using Markov Chain Monte Carlo (MCMC) with importance sampling. The Boltzmann weight, $e^{-S}$, is highly peaked, consequently analysis on $10^3$--$10^4$ importance sampled configurations 
provide accurate estimates of many interesting quantities. These configurations are characterized by five input parameters, the QCD coupling constant and four (up, down, strange and charm) quark masses\footnote{The other two quark masses, bottom and top, are too heavy to significantly affect the gauge-field distribution.}. On these configurations, correlation functions, $O$, are calculated as expectation values (ensemble averages):
\begin{equation}
\langle O \rangle = \frac{1}{Z} \int {\cal D}U \ e^{-S[U]} O[U] \ \approx \frac{1}{N} \sum_i O[U_i]
\end{equation}
where $Z= \int {\cal D}U \ e^{-S[U]}$ is the partition function, $S$ is the action of the theory that is a functional of the gauge configurations $U_i$, and $N$ is the number of importance sampled configurations, i.e., configurations 
generated with the distribution $e^{-S[U]}$. 
For example, the properties of a proton are obtained by choosing $O$ to be the 2-point function illustrated in Fig.~\ref{fig:lat_diags} (left). It represents the creation of a proton by putting in an external source consisting of quark fields with proton's quantum numbers, allowing it to evolve in time, and then annihilating it at a later time $\tau$. Its interactions with say electromagnetism are obtained by inserting an additional probe with the quantum numbers of the vector current (marked by $\otimes$) at intermediate times $t$, i.e., the 3-point function shown in Fig.~\ref{fig:lat_diags} (right). As the number of such correlation functions and their complexity have grown to provide detailed predictions of QCD, their measurement is now exceeding the cost of generation of configurations.

The expectation value of each correlation function has statistical and systematic errors. The latter is due to discretization (putting QCD on a finite lattice of dimension $(L/a)^4$ and spacing $a$). The field theory is recovered in the limit  $a \to 0$ with $L$ held fixed in physical units.  Real-world physics is obtained by tuning the coupling and quark masses to their physical values by matching predictions of five hadronic observables (for example, the masses of $\pi^+$, proton, kaon, $\Omega^-$ and $\eta_c$ hadrons) to their experimental values.  Results in the continuum limit and with physical values of the quark masses are achieved by performing simulations at multiple values of the lattice volume and spacing and then extrapolating. This is a costly procedure. To provide context, these classical simulations are analogous to those of spin models (Ising, XY, $O(N)$, etc.) except that both the generation of decorrelated configurations and measurements of 
correlation functions require about $10^9$--$10^{12}$ times more flops.  

ML methods are being developed to speed up all three steps in such calculations~\cite{Boyda:2022nmh}: (i) generation of gauge configurations (see Sec.~\ref{sec:GG}), (ii) measurement of correlation functions~\cite{Favoni:2020reg, Bulusu:2021rqz, Detmold:2021ulb, Zhang:2019qiq} (see example below), and (iii) their analysis to extract physics~\cite{Yoon:2018krb,Kades:2019wtd,Chen:2021giw}. The first two steps in LQCD, and ML systems, are computational and are therefore labeled ``black boxes''.   In Sec.~\ref{sec:bb}, we give our thoughts on why they have this label and contrast their opacities. 

Once expectation values have been calculated, they are characterized by five input parameters, and can be expanded in terms of these. Thus, in principle, one expects correlations between them, but at what level---expectation values or correlation functions?  A first such application of ML methods to predict correlation functions that are harder (more expensive) 
 to calculate using the input of simpler ones was demonstrated 
 in~\citet{Yoon:2018krb} using a GBDT Regression algorithm available in scikit-learn python ML
library~\cite{scikit-learn}.  The gain in the computational cost of expectation values was modest, 9--38\%. Using different ML models does not significantly improve 
the gain, presumably because the correlations are strong and essentially linear and the GBDT algorithm does a very good job of finding them. Nevertheless, this demonstration is non-trivial since the ML algorithm
took as input a two-point correlation function and output a set of three-point functions configuration by configuration. The ML system provided no picture or understanding of the correlations between fluctuations on a given configuration, but it did find them! We discuss this point further in Sec.~\ref{sec:bb}. 

The second important contribution of~\citet{Yoon:2018krb}, a general method for bias correction to remove the systematic errors from such ML-based predictions in expectation values, is reproduced in Sec.~\ref{sec:bias}.

\section{ML methods for efficient generation of gauge configurations: application to lattice QCD}
\label{sec:GG}

Generation of ensembles of configurations with importance sampling via Markov Chain Monte Carlo (MCMC) methods is an essential step in numerical simulations of the path-integral of statistical mechanics systems and field theories.  MCMC methods provide an efficient way to  generate samples from the highly peaked probability density functional characterizing the theory by turning it into a stationary distribution.  The limitation is that since Markov chains are stochastic processes, local~\cite{Wilson:1975id,Creutz:1984mg} or small--step-size global update schemes such as Hybrid Monte Carlo (HMC)---the algorithm of choice in today's simulations---suffer from long autocorrelation times (critical slowing down) near the quantum critical points defining the continuum limit. 

In a HMC algorithm using microcanonical time evolution, changes are proposed by generating independent Gaussian distributed momentum variables conjugate to the position variables. This system is evolved using microcanonical dynamics (it keeps the total energy fixed) with a small step size (to reduce discretization errors) but for $O(1)$ unit of time. The proposed configuration is accepted using the probability $e^{-\Delta H}$, where $\Delta H$ is the change in the action (energy)~\cite{Metropolis:1953am,Creutz:1984mg}. This evolution is guaranteed to converge to the desired Boltzmann distribution $e^{-H} \equiv e^{-S}$. At each step of the microcanonical evolution detailed balance (reversibility) is maintained, i.e., $P(A)P(B|A) = P(B)P(A|B)$ for the initial and final configurations $A$ and $B$. (If not, one can correct by the ratio of probabilities of the proposed configuration and its inverse at the accept/reject step.) It also satisfies the ergodicity requirement. However, even though the momentum variables are refreshed at the start of each trajectory, keeping the system on a constant energy surface  to enforce small \(\Delta H\) (and, thus, a large acceptance), the configuration (position variables) changes by a small amount and the evolution suffers from critical slowing down as the correlation length in the continuum limit $a \to 0$ becomes infinite in lattice units. This limit is needed to recover the field theory.  In practice, one pushes the calculations to small enough $a$ (high enough energy scale) where various perturbative expansions become reliable and can be used to extrapolate to $a = 0$. With current algorithms, and for the precision required by experiments, the generation of lattice configurations is facing the problem of large autocorrelations before this desired matching is achieved. 

It has been difficult to design efficient global update schemes that make large changes leading to fast decorrelations between configurations, as required for high statistics calculations. This is in spite of the few requirements on proposed changes as stated above: ergodicity (all configurations have equal weight, any configuration should be reachable), and detailed balance (reversibility) is maintained so that $P(A)P(B|A) = P(B)P(A|B)$ for any two configurations $A$ and $B$. After thermalization, the configurations should satisfy the Gibbs distribution. What has revolutionized the calculations over the last decade is the development of the adaptive multigrid algorithm~\cite{Brannick:2007ue} for inverting the Dirac operator, the essential but extremely time-consuming part of the microcanonical evolution. The goal is to replace the proposed change based on microcanonical evolution with a ML algorithm, especially those that can be trained without generating a sufficiently large ensemble in the first place. 

Here we outline the promising method of normalizing flows~\cite{Kobyzev:2021} for the generation of gauge configurations that is being actively pursued. The goal is to combine ML with the MC method to generate configurations without autocorrelations. 

\begin{figure}[tb]
    \centering
    \includegraphics[width=0.95\textwidth]{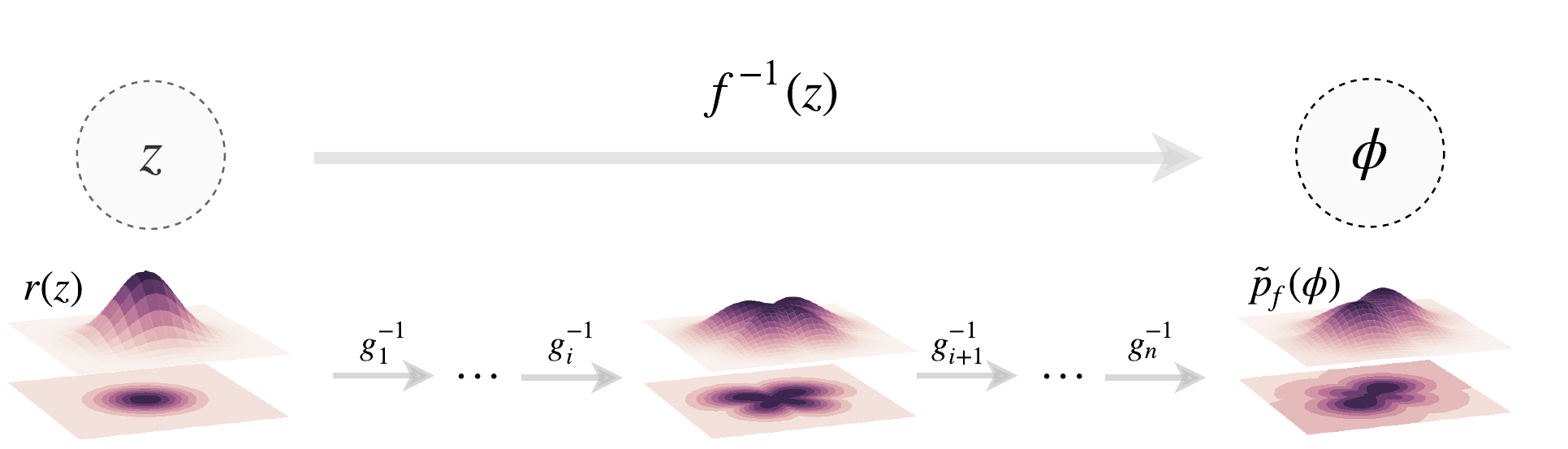}
    \caption{Normalizing flow transforming the samples $z$ from a simple distribution $r(z)$ to the samples $\phi$ that follow the distribution $\tilde{p}_f(\phi)$, an approximation to ${p}_f(\phi)\propto e^{-H}$, 
    through the mapping $\phi = f^{-1}(z)$ constructed by stacking the inverse layers $g_i^{-1}$. The figure is taken from \citet{Albergo:2019eim}.
    }
    \label{fig:flow}
\end{figure}

Mathematically, there exist transformations that take a lattice configuration of uncorrelated variables (independent Gaussian distributed variables) to any desired Boltzmann ensemble. The practical question is: can machine learning be used to find approximations to such transformations that (i) preserve the symmetries (local gauge invariance, parity, etc) and conservation laws (Gauss's law) of the theory, (ii) maintain ergodicity,  and (iii) have a tractable Jacobian that is needed to calculate the probability of the proposals? Finding such a transformation allows one to propose large global changes (since the proposed new configuration is independent of the old one) in a Metropolis accept/reject scheme~\cite{Metropolis:1953am,Creutz:1984mg}.  
If the approximation is good enough to make the acceptance rate reasonable, using ML to transform Gaussian variables to the normalized distribution  will generate an ensemble at a much lower cost and with essentially no autocorrelations. Correctness (generating configurations with the desired Boltzmann distribution) will be preserved provided the ML process is reversible and ergodic. For simple actions, the efficacy of the neural network based normalizing flow, illustrated in Fig.~\ref{fig:flow}, has already been demonstrated, see \citet{albergo2021introduction} for a review. There is considerable activity in the community to extend the method to QCD,  a non-abelian gauge theory in four dimensions~\cite{Boyda:2022nmh}.

\section{Contrasting two black boxes: Lattice QCD and ML}
\label{sec:bb}

In this section, we present some thoughts on statements commonly made about both LQCD and ML---that they do not provide intuition or an understanding of the calculation and are essentially black boxes, and therefore unappetizing.

LQCD involves two steps that are ``black\rlap.''{\spacefactor\sfcode`\.} The first is the generation of ensembles of
gauge configurations.  These configurations provide a statistical description of the ground (vacuum) state of QCD.  One has no quantitative measure or understanding 
of what the 
fluctuations in any of the $SU(3)$ matrices attached to each link on the lattice mean, not only because of local gauge invariance, but also because the examination of a single finite volume configuration does not specify the distribution from which it is drawn. The second is the construction of the quantum 
wavefunction of a state and its properties. For example, 
the mass of the proton, $M$, is obtained from a fit to the expectation value of the 2-point function, $C_{\rm 2pt}$, versus the separation $\tau$ between the source and the sink points
in Euclidean time, $C_{\rm 2pt} \sim e^{-M \tau}$.  The building block of $C_{\rm 2pt}$ is the
quark propagator, which is the inverse of the Dirac operator defined on a given configuration. It has no knowledge of a proton or any other hadron or any $n$-point correlation function.  Information of the proton is built on averaging this 2-point
function over thousands of configurations (doing the path integral) that picks out the relevant gauge interactions---it constructively adds fluctuations in gauge fields to create the wavefunction that correlates with the quantum numbers of the created state. It does this at each time $\tau$ and also builds into the wavefunction the knowledge that as $\tau$ increases, the excited states are exponentially damped. Said another way, it is the averaging  that specifies the distribution. 

Unfortunately, the simulation process gives us no idea of what the gauge fields or their fluctuations that exist within a proton state (i.e., its wavefunction) look like.
The computer constructs the fully quantum wavefunction through the averaging process but does not provide 
any picture of it in terms of the spatial distribution of gauge fields.  But once the wavefunction is created ``within the computer'', all gauge invariant quantities such as the distribution of 
the electric charge, spin, momentum of quarks within the proton can be calculated. In summary, because the vacuum state and the wavefunction of any state created via appropriate external probes has no analytical or pictorial representation, simulations of QCD are regarded as a black box.  

The calculated expectation values of correlations functions are well-behaved numbers that are rigorous outputs of QCD.
They are trusted not only because of theoretical arguments, but also because the hundreds of results one obtains from the same
set of configurations by measuring different correlation functions
exhibit all the subtle features of QCD. And all results obtained so far agree with experimental measurements where available~\cite{Aoki:2021kgd}. Their variations with respect to the five tunable input parameters are also understood and quantifiable as power series\footnote{It is implicit that we first need to isolate nonanalytic terms known from general arguments.} with approximately known coefficients, especially near the physical point where the artifacts are small and perturbative expansions in $a$, the pion mass $M_\pi$ (surrogate for masses of the light, $u$ and $d$ quarks) and $L$ work. Fits using these ans\"atze are then used to extrapolate data to physical values of the input parameters. Bias (systematic uncertainty) in results can arise from an incomplete sampling of the phase space of the path integral, usually checked by increasing the 
statistical sample by factors of 10--100, and those due to using truncated versions of the 
fit ans\"atze for the extrapolations by increasing the number of data points and simulating closer to the physical point in a controlled way. 

ML has the potential to provide the ensemble of decorrelated lattices much more cheaply, to predict an expectation value with the input of another~\cite{Yoon:2018krb}, and to interpolate and or extrapolate results to other values of the input parameters~\cite{Boyda:2022nmh}. It too is regarded as a black box for the following reasons.

For many applications, the ML model 
can be viewed as finding a ``spline'' to parameterize the 
data within a certain range of parameter values for an unknown function or a function that is very expensive to calculate. It is a black box because the ``spline'' is 
the very large number of weights on links in a network that have no physical or intuitive connection to the problem (see developments in Sec.~\ref{sec:AdS}). Furthermore, the number of these links is much larger than 
the degrees of freedom in the data. The process of getting a good ML model 
consists of (i) choosing the [neural] network 
architecture, (ii) the training schedule and the cost
function, (iii) stopping criteria for the training, and (iv) the choice of 
data representation. There usually is little or no quantifiable 
determinism or logic for choosing between options for any of these four steps. 
The practice is to develop a model, train it and see if it works on a given 
data set and then tweak these steps. Even when a model works, 
the connectivity and the weights remain a black box.

Bias in results, from ML systems where the desired distribution cannot be guaranteed, can come from any of the above four steps, and if 
the data set is not comprehensive. For example, if the 
data are limited or there was a bias in the collection or the training set is 
insufficient or different from the test set. 
Having a comprehensive data (and training) set is essential. 
Unfortunately, when one is 
looking for a tiny or uncharacterized signal, there is no, a priori,
way to know if the training set is comprehensive other than a failure. 
One practical method for exposing failures, biases and lack of robustness is continuous monitoring of results to look for the lack of robustness under variations in the four steps and adding more data at all levels, something unappetizing to purists.  

Preserving the symmetries of QCD in LQCD formulations and calculations play a 
very important role. In ML too, one can impose symmetries at the level of the 
model (a penalty for models that break it or build it into the tuning of weights) or 
data (data representation itself builds in the symmetry or perform data augmentation 
to realize the symmetry). Ongoing work suggests that incorporating 
an understanding of the science or 
the dynamics of the system into ML is likely to dramatically increase its power \cite{Favoni:2020reg, Bulusu:2021rqz, Kanwar:2020xzo, Maitre:2021uaa}. 

Overall, Lattice QCD is a rigorous field theory but its simulations are a black box. To merge it with 
ML, one should simply continue to regard simulations of it as a computational challenge---start with input values of the five parameters and  predict the correlation functions, or even the expectation values directly. The guts of the process can be an efficient combination of simulations of lattice QCD and ML. 

The intuitively apparent difference between ML and simulations of lattice QCD is that in LQCD the same set of gauge configurations provide the answers to all physics questions that can be formulated in terms of correlation functions whereas in ML, one expects a different 
setup with respect to any of the four steps may be more appropriate for each observable or 
application as even the underlying notions of the vacuum state and the  wavefunction are lost. We re-emphasize this point: the fluctuations in LQCD configurations contain information on all correlation functions, even correlation functions that violate the symmetries of QCD, for example, parity, charge conjugation, or baryon number. All one needs in the latter case is to calculate the appropriate correlation functions and the reweighting factor to account for the difference in the Boltzmann weights. Naively, one has a hard time accepting that such richness could be built into a ML system trained on one set of correlation functions. Examples of progress in this direction are given in Sec.~\ref{sec:aware}. Thus, there is a strong possibility that even this difference will blur with experience. 

Our hunch is that the generation of decorrelated configurations in LQCD using methods such as normalizing flows~\cite{Kobyzev:2021} will be one of the first successes beyond the examples of predicting expensive to calculate correlation functions in terms of cheaper ones as discussed in~\citet{Yoon:2018krb}. Another exciting application of ML in the analysis of data is to estimate the real-time spectral function from lattice data for the Euclidean 2-point function~\cite{Kades:2019wtd,Spriggs:2021dsb,Horak:2021syv,Chen:2021giw}, a notoriously difficult problem since the latter is the Laplace transform of the former. 

As these ML methods mature, a key issue will be the detection and correction of bias in ML predictions. The next section first describes a general method for bias correction followed by some additional comments. 
\section{A general method for bias correction in ML predictions of averages}
\label{sec:bias}

Consider $M$ samples of independent measurements
of a set of observables $\mathbf{X}_i = \{o_i^1, o_i^2, o_i^3,
\ldots\}$, ${i=1,\ldots,M}$, but the target observable $O_i$
is available only on $N << M$ of these.
These $N$ are called the \emph{labeled data} ($LD$) and the remaining $M-N$ are 
called the \emph{unlabeled data} ($UD$). The goal is
to build a ML model $F$ that predicts the target observable
${O_i\approx O_i^{\textrm{P}}\equiv F(\mathbf{X}_i)}$ by
training a ML algorithm on a subset $N_t<N$ of the labeled
data. The bias corrected estimate \(\overline
O\) of \(\langle O\rangle\) is then obtained as\looseness-1
\begin{align}
\overline{O} =  \frac{1}{M-N} \sum_{i\in \{UD\}} O^\textrm{P}_i 
+ \frac{1}{N_b} \sum_{i \in \{BC\}} (O_i - O^\textrm{P}_i)\,,
\label{eq:o_unbiased}
\end{align}
where the second sum is over the ${N_{b}\equiv N-N_t}$ remaining
labeled samples that corrects for possible bias. Here
\(O^\textrm{P}_i\) depends explicitly on \(\mathbf{X}_i\) and
implicitly on \(N_t\) and all training data \(\{O_j,
\mathbf{X}_j\}\). For fixed ML model $F$, the sampling variance of
$\overline{O}$ is then given by
\begin{align}
    \sigma_{\overline O}^2 = \frac{\sigma_O^2}N \left\{s^2\frac N{M-N}+ \frac1f [(1-s)^2 + 2s(1-r)]\right\}\,,
\label{eq:var}
\end{align}
where \(\sigma_O^2\) is the variance of \(O_i\),
\(s\equiv\sigma_{O^P}/\sigma_O\) is the ratio of the standard
deviations of the predictor variable \(O^P\) to the true observable
\(O\), \(r\) is the correlation coefficient between these two, and
\(f\equiv N_b/N\) is the fraction of observations held out for bias
correction.  Eq.~\eqref{eq:var} shows that when \(s\approx1\approx
r\), this procedure increases the effective sample size from \(N\),
where \(O_i\) are available, to about \(M-N\). For simplicity, in
deriving Eq.~\eqref{eq:var}  details such as the
statistical independence of the data was ignored.
One way to account for
the full error, including the sampling variance of the training and 
the bias correction data sets, is by using a bootstrap
procedure~\cite{10.2307/2958830} that independently selects \(N\)
labeled and \(M-N\) unlabeled items for each bootstrap sample.
While the
bias correction removes the systematic shift in the prediction, it can
increase the final error, i.e., the systematic error due to biased ML prediction is converted
to a statistical error–––the advantage is, of course, that the standard statistical techniques like bootstrap can estimate the latter accurately. Nevertheless, since the bias is explicitly estimated in this procedure, in specific applications one can evaluate whether to prefer a precise biased estimator to an unbiased, but imprecise, estimator.

%

\section{Examples of physics-aware machine learning models}
\label{sec:aware}

\begin{figure}[tb]
    \centering
    \includegraphics[width=0.95\textwidth]{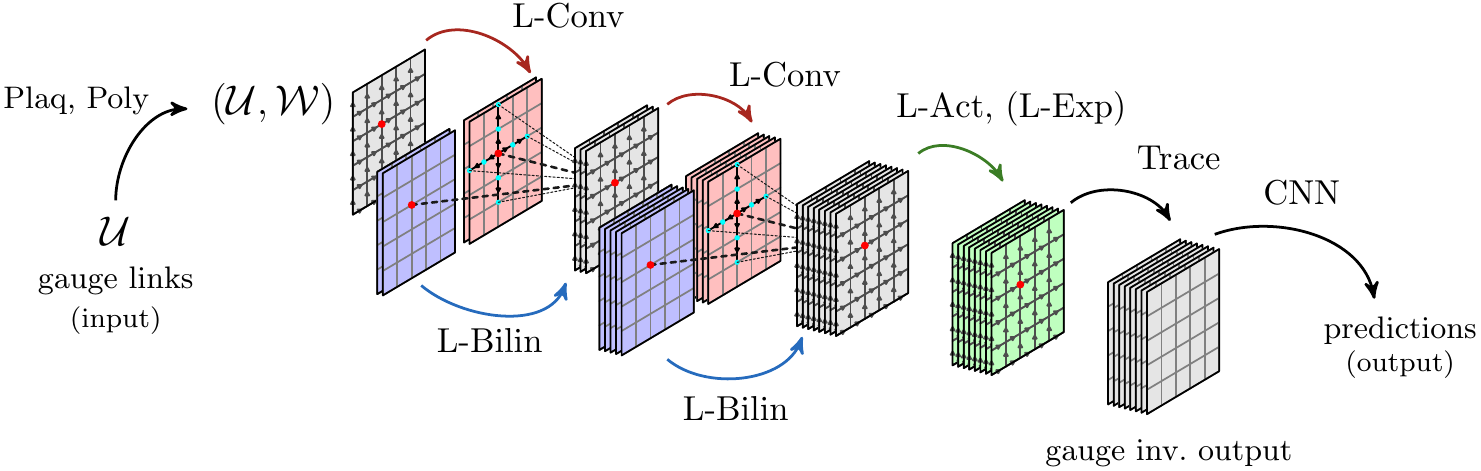}
    \caption{An example of the gauge equivariant L-CNNs constructed from the lattice gauge equivariant convolutions (L-Conv), bilinear layers (L-Bilin), activation functions (L-Act), exponentiation layers (L-Exp), trace layers (Trace), and plaquette and Polyakov layers (Plaq and Poly). The input of lattice gauge fields, $\cal{U}$, and Wilson lines ending at the starting point, $\cal W$,  is processed by the operations preserving the physical symmetry of the gauge equivariance to make accurate predictions. The figure is taken from \citet{Favoni:2020reg}.
    }
    \label{fig:lcnn}
\end{figure}

When applied to physics data, ML learns about the symmetries and constraints of the system inferred from the data and utilizes such information to make predictions. By explicitly implementing the physical rules in the ML procedure, however, one can obtain more precise ML predictions with a smaller number of training data. In \citet{Favoni:2020reg}, for example, the authors proposed a new structure of the convolutional neural network, named Lattice Gauge Equivariant Convolutional Neural Networks (L-CNNs), that preserves gauge equivariance. They introduced a set of network layers that preserve gauge symmetry exactly: convolutions, bilinear layers, activation functions,  exponentiation layers, trace, plaquette layers, and Polyakov layers, which can be stacked arbitrarily to form gauge equivariant networks. An example of the L-CNNs construction is illustrated in Fig.~\ref{fig:lcnn}. Also studied is how to impose global translational invariance on convolutional neural network (CNN) architectures for the analysis of scalar field configurations on two-dimensional lattices~\cite{Bulusu:2021rqz}. For the study of the gauge generation, described in Section~\ref{sec:GG}, a flow-based generative model satisfying the exact gauge invariance for lattice gauge theories has been proposed by defining gauge-invariant flows and coupling layers~\cite{Kanwar:2020xzo}. Also, in \citet{Maitre:2021uaa}, the accuracy of the ML predictions of the matrix elements for $e^+e^- \rightarrow$~up~to~5-jets, described in Section~\ref{sec:amp} has been significantly improved by exploiting the knowledge of the factorization properties of the matrix elements.

When constrained by an incorrect theory, however, the trained ML tends to generate biased predictions. In Eq.~\eqref{eq:pdf}, for example, a prefactor $x^{1-\alpha_k} (1-x)^{\beta_k}$ multiplies the neural network parametrization of the PDFs to enforce the large-$x$ behavior $f_k(x=1)=0$ required by the constituent counting rules~\cite{PhysRevLett.31.1153} and the small-$x$ behavior inspired by the Regge theory~\cite{PhysRevLett.22.500}. However, the prefactor provides more strict constraints than those allowed by the physics rules and may introduce bias in the predictions and uncertainty estimations~\cite{Carrazza:2021yrg}. To avoid such bias, \citet{Carrazza:2021yrg} proposed to implement minimal constraints by removing the prefactor and replacing Eq.~\eqref{eq:pdf} by 
\begin{align}
    x f_k(x, Q_0) = A_k~\big[\textrm{NN}_k(x) - \textrm{NN}_k(1)\big]\,.
    \label{eq:pdf2}
\end{align}
They find no loss in efficiency and good agreement with previous results. 

\section{Unsupervised ML approaches}
\label{sec:unsupervised}

Modern physics experiments and simulations produce a deluge of data but many analysis models are not capable of dealing with and getting the most out of such large and complex data. However, the complicated physics data can be explained by a small number of features lying on a lower-dimensional manifold. For example, translational symmetry of a physics system could explain all the data symmetric under translations. Unsupervised ML is a class of algorithms drawing inferences from untagged data by building a compact internal representation of the data. By applying unsupervised ML algorithms to physics data, one can find a natural clustering and key features of the data, find anomalous events, and generate synthetic data. 

\begin{figure}[tb]
    \centering
    \includegraphics[width=0.95\textwidth]{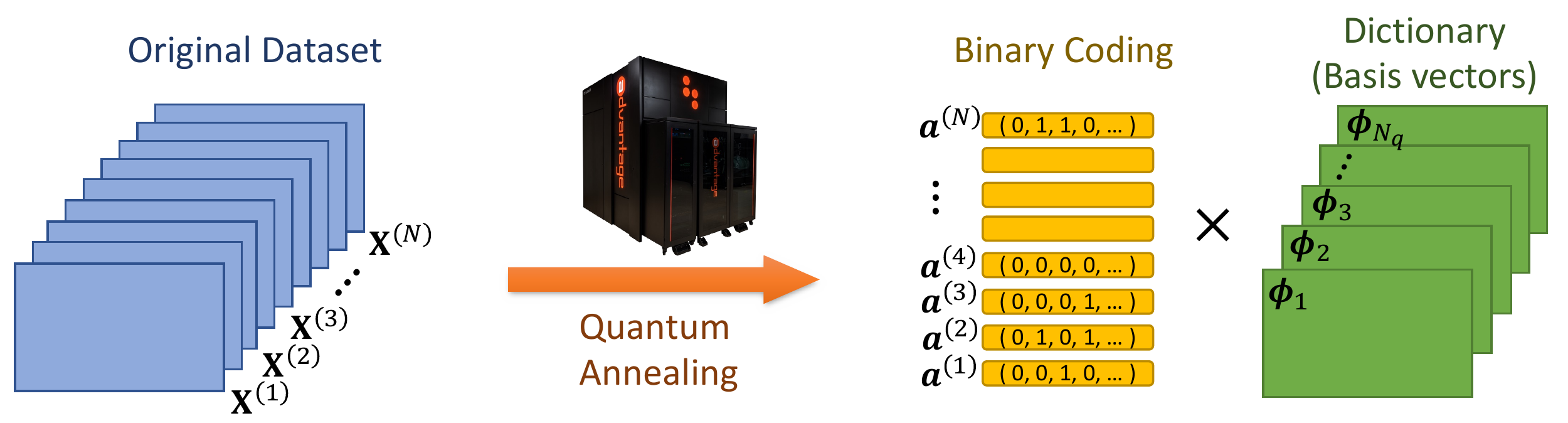}
    \caption{An illustration of the ML-based compression algorithm proposed by \citet{Yoon:2021btl}. Exploiting the correlation in physics data, the algorithm finds a lower-dimensional binary representation of the original floating-point data that can be stored in a small memory space but can precisely reconstruct the original data from a set of basis vectors. A quantum annealer was used to solve the NP-hard optimization problem of finding the optimal binary coding. 
    }
    \label{fig:bin_comp}
\end{figure}

Although experimental data is one of the best candidates for the unsupervised ML applications (for example as discussed in Sec.~\ref{sec:event} for classification of events, and see the chapter by David Rousseau, {\it ibid.}), many studies find useful applications in the analysis of data from theoretical particle physics. The generative models used for the Monte Carlo simulations explained in Section~\ref{sec:integ} are good examples of such applications. \citet{Carrazza:2021hny} also applied generative adversarial networks to produce pseudo-samples of the Monte Carlo PDF data. With the dataset enhanced by the synthetic replicas, they were able to find a better subset accurately representing the underlying probability distribution of the original data set than the standard method. Exploiting the correlation between physics data, \citet{Yoon:2021btl} developed a lossy compression algorithm for statistical floating-point data through a representation learning with binary variables on quantum annealers, as illustrated in Fig.~\ref{fig:bin_comp}. Based on the algorithm explained in Section~\ref{sec:bias}, they further presented a bias correction method for the inexact reconstruction from lossy-compressed data.

\section{Exploring the space of possible theories}
\label{sec:BSM}

Even though the standard model of particle physics (SM) is an incredibly successful theory with no known deviation between its predictions and any known experiments, it, nevertheless, has major shortcomings. Three important ones are related to the large scale structure of the observed universe: dark matter, dark energy, and an explanation for the matter-antimatter asymmetry. Possible mechanisms for the latter are baryogenesis or leptogenesis~\cite{Sakharov:1967dj,Barrow:1981bv,Fukugita:1986hr,Davidson:2008bu}. Thus the SM 
needs to be extended to include the standard model of cosmology and possibly particles and interactions from beyond the standard model needed to explain these observations.  At an even more fundamental level, the classical theory of general relativity that explains gravity and the large scale evolution of the universe has not been successfully quantized. A fully unified theory needs to include quantum gravity. The number of possible theories are, however, huge. Given that the search for the fundamental or the ``next'' standard model is at the core of HEP research, it is not surprising that ML methods are already being used to help with this construction. 

We discuss two examples. The first is to make predictions of a given BSM theory defined by specific values of the unknown parameters, particles and interactions, and generate synthetic data with a small number of output variables that characterize its predictions as described in Sec.~\ref{sec:amp}. To determine if it is the theory of nature, one would again design a second ML system to compare predictions to  experimental data.  A flow chart of calculations to efficiently explore the space of input parameters (using the minimal supersymmetric model (MSSM) theory as an example) is as follows. Calculate the NLO cross-sections using conventional approaches for well-distributed values of the MSSM model parameters, such as gluino and squark masses,  and interpolate these results using DGP to  obtain the cross-sections for arbitrary values of model parameters. Again, symmetries, conservation laws, and other physics knowledge can be built into ML models or the output. The space of allowed couplings can be progressively constrained by matching these predictions to experimental data. Such ML systems can then be extended to search between BSM theories. 

\begin{figure}[t]
    \centering
    \includegraphics[width=0.5\hsize]{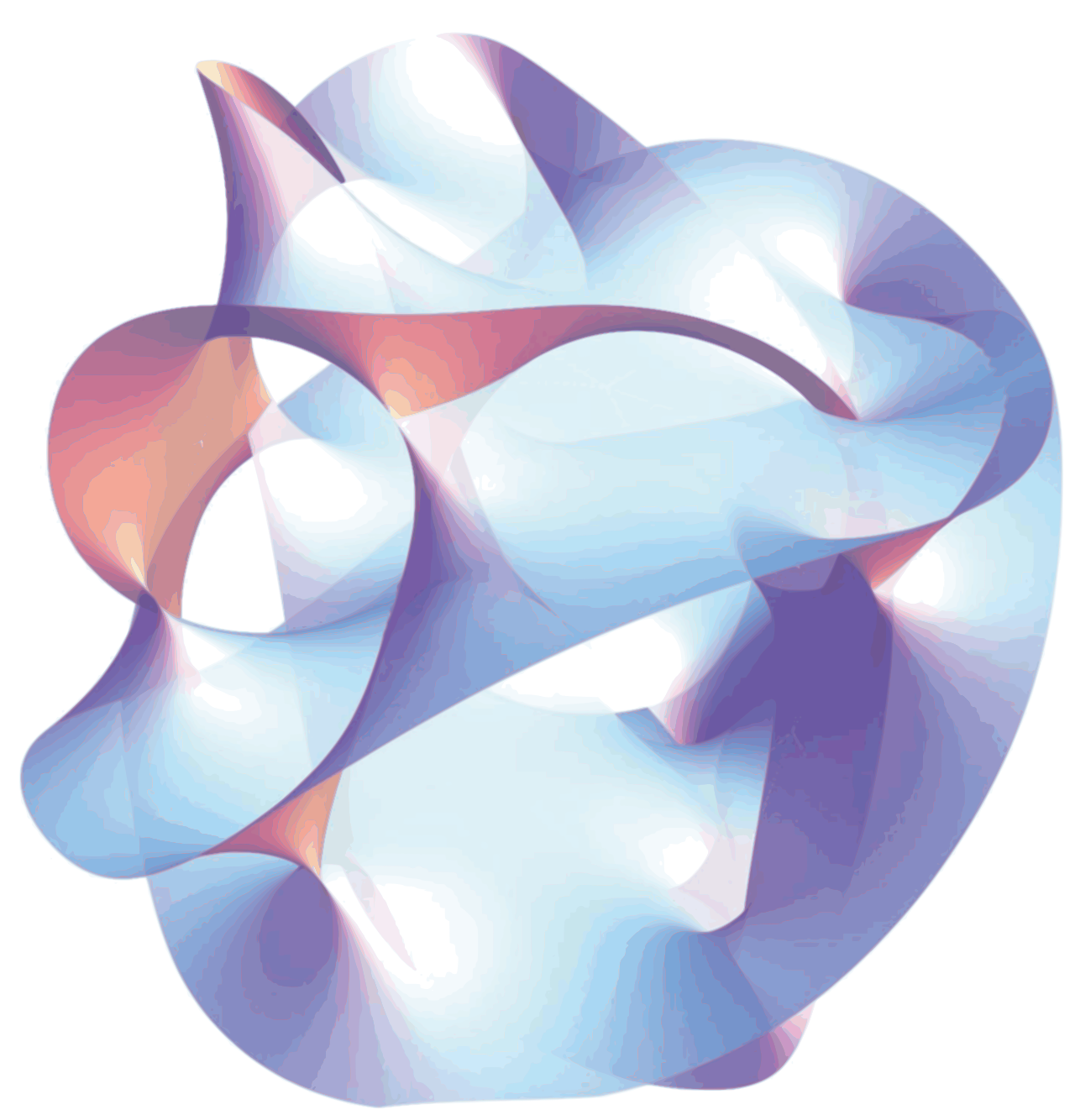}
    \caption{A cross-section of the quintic Calabi-Yau manifold.  In the F-theory approach to string theory, the various string vacua arise from choosing a Calabi-Yau manifold structure and values of fluxes on it. Picture in public domain~\cite{wikimediafigure} where it was creared using the methods described in~\citet{Hanson94aconstruction}.}
    \label{fig:Fvacua}
\end{figure}

The second example is string theories. The reigning fundamental theory that incorporates both general relativity and quantum field theory is called the `M-theory'~\cite{Witten:1995ex}. It is believed to be unique and various string theories appear as fluctuations  around its many vacuua or false vacuua. The only freedom one has in developing a model for our universe is choosing between these vacuua. Unfortunately, these vacuua are parameterized by a set of integers, and by some estimates, there are more than \(10^{10^5}\) of these~\cite{Taylor:2015xtz}. A visualization is shown in Fig.~\ref{fig:Fvacua}. This poses three grand challenge questions: how to choose which vacuua to analyze, do \emph{any} of these have the standard model as a low energy effective theory, and what are the novel predictions of these theories~\cite{Svacua,HE2017564}.

At their core, these questions can be framed as an algebraic problem.  For example, the number of `particle species' that exist around any of these vacuua is an integer-valued function of the integer parameters that characterize the vacuum.  This is a very difficult function to evaluate, and ML has been used to get surrogate functions for problems like this with over 90\% match to the true function in some domains~\cite{Ruehle:2017mzq,He:2017aed}.  More interestingly, one can train an ML architecture to identify regions of parameter space within which polynomial approximations with theoretically expected structures work, or where a certain structure believed to be important for the standard model or our universe exists~\cite{parr2020machine,Deen:2020dlf}.

This example illustrates the strength of combining human ingenuity in formulating string theory and elucidating the landscape of possible vacuua  with ML tools to sift through the very large number of these vacuua!

\section{Interpreting extra dimensions from deep neural nets}
\label{sec:AdS}

Lastly, a particularly interesting connection of deep neural networks to modern theories of HEP is provided by the recently discovered latticized AdS-CFT correspondence~\cite{Hashimoto:2018ftp,Hashimoto:2019bih}. In 1998, Maldecena made the profound discovery that conformal field theories (CFT) appear as the description of the physics at the remote boundary of gravitational theories on an anti-deSitter (AdS) background with a black hole horizon at its inner boundary, i.e., the AdS/CFT correspondance~\cite{Maldacena:1997re,Ramallo:2013bua}.  From the point of view of the CFT, the meaning of the extra dimension is completely opaque: but~\citet{Hashimoto:2018ftp} and~\citet{Hashimoto:2019bih} provide an interpretation of the slices along the extra dimension as the layers of a neural network that computes the black hole horizon from the field theory data. Specifically, a neural net was designed such that its internal weights, interpreted as metric components on a discretized space-time, satisfy the gravitational equations of motion.  Since the black hole boundary is fixed and known, one can now `train' this network to `learn' the black-hole boundary condition when shown simple known field conformation as input at the remote boundary, and the weights in the trained network would then approximate the AdS geometry corresponding to the boundary theory. 

\section{Conclusions}
\label{sec:conclusions}

Hard computational problems abound in theoretical high energy physics. Dealing with the complexity of each event at the high luminosity LHC and the rate of events, sub-percent precision from lattice QCD, and finding the right vacuum of M-theory that describes our universe all require new tools.  This chapter touched on how machine learning methods for optimization, pattern recognition (e.g., collider events), numerical integration, generative models, extrapolation, interpolation, regression, and surrogate models can transform many areas of research.  

We have devoted considerable space to lattice QCD, not only because  ML methods hold the promise to significantly reduce the large computational needs but because LQCD illustrates how, by giving up analytical control (understanding, visualization) in the intermediate steps of calculations, the rich structure and predictions of strongly interacting quantum field theories can be obtained. Relying on large-scale simulations of lattice QCD and turning QCD into a numerical problem may not satisfy a purist but results from simulations have become palatable because of the rigorous foundation of the calculations in the Feynman path integral formulation. What these simulations highlight is that while the creation of the quantum ground state and the wavefunctions of hadrons in the computer is a ``black box'', this statistical description provides the means to calculate many observables of interest with increasing precision. In the future, methods such as normalizing flows are likely to supplant standard Markov Chain Monte Carlo methods for generating gauge configurations---all that is needed is a transformation, no matter how opaque, from a simple distribution to the desired distribution. The evolution in the thinking of a theorist needed is that it does not matter how one gets, say an ensemble of configurations for lattice QCD, as long we can validate they obey the Boltzmann distribution and guarantee that the results for correlation functions are bias-free. \looseness-1

We anticipate that, even as ML methods become more powerful and calculations using them become opaque, the need for understanding the physics will not decrease, and physics-aware development of ML will be more efficient. The trend towards collaborative teams of physicists, mathematicians, computer scientists, and statisticians is bound to enrich research, especially in areas requiring hard calculations that need to be repeated billions of times (event generators, lattice QCD) or to manage and process large complex data sets (experiments at the LHC).

\section*{Acknowledgments}
We thank Phiala Shanahan for reading though the manuscript and providing feedback. T. Bhattacharya and R. Gupta were partly
supported by the U.S. Department of Energy, Office of Science, Office
of High Energy Physics under Contract No.~89233218CNA000001. 
T. Bhattacharya, R. Gupta, and B.Yoon were partly
supported by the LANL LDRD program.
\newpage 
\bibliographystyle{IEEEtran}
\bibliography{refs} 

\end{document}